\documentclass[aps,pra,reprint,twocolumn,amsfonts,amssymb,amsmath,floatfix,superscriptaddress,tightenlines]{revtex4-1}

\allowdisplaybreaks
\usepackage{graphicx}
\usepackage{pdfpages}
\usepackage{relsize}
\usepackage{bm,times,dsfont}
\usepackage[colorlinks=true,citecolor=blue,urlcolor=blue]{hyperref}
\numberwithin{equation}{section}

\makeatletter
\AtBeginDocument{\let\LS@rot\@undefined}
\makeatother

\begin{document}
\title{Restoring gauge invariance in conventional fluctuation corrections to a superconductor}

\author{Rufus Boyack}
\affiliation{Department of Physics \& Theoretical Physics Institute, University of Alberta, Edmonton, Alberta T6G 2E1, Canada}

\begin{abstract}
The complete set of diagrams for conventional fluctuation corrections to the normal state of an ultraclean superconductor is derived using the functional-integral approach. 
The standard diagrams which characterize the fluctuation phase of a superconductor, namely, Aslamazov-Larkin, Maki-Thompson, 
and Density of States, are obtained and proved to be insufficient to produce a gauge-invariant electromagnetic response. 
An additional diagram is derived, and it is found to be essential for resolving a problematic and underappreciated aspect of the conventional fluctuation formalism. 
It is only with the inclusion of this additional diagram that the theory is now rendered gauge invariant and does not display a normal-state Meissner effect. 
It is shown that not only the Aslamazov-Larkin term but all of the microscopic diagrams are encapsulated within a Gaussian-level treatment of the effective action. 
\end{abstract}

\maketitle

\section{Introduction}

Excitement surrounding the discovery of high-$T_{c}$ materials has led to a renaissance in the study of superconducting fluctuation phenomena~\cite{Varlamov_2018,VarlamovBook,BennemannBook,Loktev_2001}.
Indeed, the short coherence lengths and quasi two-dimensionality of these systems lead to a situation in which fluctuations play a prominent role~\cite{Keimer_2015}.
Studies~\cite{Serbyn_2009,Levchenko_2011,Varlamov_2018} based on the original formulation of superconductor fluctuations (or variants thereof) 
have enabled a good preliminary understanding of many of the electromagnetic (EM)  and thermodynamic properties of the cuprates to be achieved. 
Underlying any such study of EM transport is the fundamental tenet of gauge invariance.
Furthermore, it is imperative to ensure that there is no Meissner effect in the normal-state response of superconductors. 

The seminal papers on superconductor fluctuations by Maki~\cite{Maki_1968} and Aslamazov and Larkin~\cite{Aslamazov_1968,*KhalatnikovBook}
each provided unique insights into understanding the fluctuation conductivity of a superconductor. 
Maki's paper considered only the Maki-Thompson (MT) and Density of States (DOS) diagrams, but missed the Aslamazov-Larkin (AL) diagram, while the paper of Aslamazov and Larkin, 
which included all of the above diagrams, did not appreciate the importance of the anomalous Maki-Thompson contribution to electrical conductivity. 
Thompson~\cite{Thompson_1970} further elucidated this importance in two-dimensional (2D) superconductors where he showed that, 
without proper regularization, the anomalous MT conductivity is divergent~\cite{VarlamovBook}. 

In this paper the functional-integral approach~\cite{Popov} is used to obtain the complete EM response for the fluctuation phase above $T_{c}$, and resolve a problematic aspect of the fluctuation literature. 
The issue of concern is not widely appreciated and relates to the fundamental principles discussed earlier. 
In both of their articles Maki and Aslamazov and Larkin failed to unambiguously establish that their fluctuation formalisms are gauge invariant. 
In fact, other important fluctuation literature~\cite{Aslamazov_1980,Aslamazov2_1980} claimed that, with only the AL, MT, and DOS diagrams, there was no anomalous diamagnetism (normal-state Meissner effect).
Furthermore, the most influential monograph~\cite{VarlamovBook} in this field also claims that ``this cancellation [between AL, MT, and DOS in the zero momentum, zero frequency limit] confirms the absence of anomalous diamagnetism above the critical temperature" (see page 141). As will be shown in this article, the aforementioned statement is incorrect. 
Here, the AL, MT, and DOS diagrams are all derived,
along with an additional diagram not previously noted in the original fluctuation literature. 
This diagram, which we refer to as a ``Gaussian-level diamagnetic diagram", is vital for establishing gauge invariance and the absence of the normal-state Meissner effect. 
In addition, it has important physical consequences as it establishes the nature of the fluctuation diamagnetism and ultraclean conductivity. 
This manuscript thus provides a complete account of the ultraclean fluctuation theory, remedies this seemingly unknown but serious error in this literature, 
and provides a link between the functional and diagrammatic methods of fluctuation phenomena. 

The Gaussian-level diamagnetic diagram is independent of external momentum, and so in that sense it bears resemblance to the diamagnetic term in the EM response of a free gas. 
Here it is derived at the Gaussian level, and hence the basis for its nomenclature. 
It is well-known~\cite{VarlamovBook} that the AL diagram can be obtained from Ginzburg-Landau theory; 
however, the MT and DOS diagrams are conventionally found from a microscopic procedure~\cite{VarlamovBook} which obscures their similarity. 
Here it is shown that MT, DOS, and the Gaussian-level diamagnetic diagram all arise from the same term in the effective-action response, 
a result which is obfuscated in the conventional approach~\cite{VarlamovBook} to deriving the MT and DOS contributions~
\footnote{To be clear, the complete ultraclean fluctuation theory comprises five diagrams: one AL, one MT, two DOS, and the Gaussian-level diamagnetic diagram. In the EM response there are two terms, one which gives rise to the AL diagram, and another which gives rise to the other four diagrams.}.
Furthermore, it is proved that the complete ultraclean theory, now with five diagrams (one AL, one MT, two DOS, and the Gaussian-level diamagnetic diagram) is 
(i) gauge invariant, (ii) has no normal-state Meissner effect, and (iii) satisfies the conductivity sum rule. 

The initial functional-based approaches~\cite{Svidzinskii_1971,SvidzinskiiBook} to studying fluctuation phenomena derived only the AL contribution.
More recent treatments have gone beyond the pioneering work of Svidzinskii and incorporated varying degrees of additional complexity. 
In this regard, in Ref.~\cite{Taylor_2006} the superfluid density was calculated at the Gaussian level, and it was noted that there is an AL contribution.
The full gauge-invariant response at the Gaussian level was derived in Ref.~\cite{Anderson_2016}, 
and it was shown how to correctly treat collective modes by consistently solving the saddle-point condition. 
In Ref.~\cite{Fanfarillo_2012} the diamagnetic susceptibility and paraconductivity for the AL diagram were derived, 
with the addition of incorporating a multichannel Hubbard-Stratonovich decomposition.
Finally, we note that by using the fluctuation formalism the inclusion of disordered interactions~\cite{Finkelstein_2012,Fiegelman_2018,Skvortsov_2018}, for both strong and weak disorder strengths, has also successfully been accomplished. 

The structure of the paper is as follows: Sec.~(\ref{sec:Fluctuation_Action}) introduces the Hubbard-Stratonovich transformation and constructs the Gaussian-level fluctuation action. 
The EM response is then computed in Sec.~({\ref{sec:EM_response}) and all the fluctuation diagrams are derived, with  
the proofs of both gauge invariance and the absence of the Meissner effect also given. The electrical conductivity is discussed in Sec.~({\ref{sec:Conductivity}), and
finally, in Sec.~(\ref{sec:Conclusion}), the conclusion is presented.

\section{Fluctuation action}
\label{sec:Fluctuation_Action}

\subsection{Hubbard-Stratonovich transformation}

The starting point is the Hamiltonian for a Fermi superfluid interacting through an attractive interaction~\cite{Altland_Simons}:
\begin{eqnarray}
H & = & \int d\mathbf{r}\sum_{\sigma}c_{\sigma}^{\dagger}(\mathbf{r})\left(\frac{\hat{\mathbf{p}}^{2}}{2m}-\mu\right)c_{\sigma}(\mathbf{r})\nonumber\\
&\quad&-\int d\mathbf{r}d\mathbf{r}^{\prime}g(\mathbf{r}-\mathbf{r}^{\prime})c_{\uparrow}^{\dagger}(\mathbf{r})c_{\downarrow}^{\dagger}(\mathbf{r}^{\prime})c_{\downarrow}(\mathbf{r}^{\prime})c_{\uparrow}(\mathbf{r}).
\end{eqnarray}
Here $m$, $\mu$ are the fermion mass and chemical potential, respectively. 
For the $s$-wave case, which is of primary interest
here, the coupling constant is $g(\mathbf{r}-\mathbf{r}^{\prime})=g\delta(\mathbf{r}-\mathbf{r}^{\prime})$,
where $g>0$ is a constant. The coherent-state functional integral~\cite{Altland_Simons}
is then $\mathcal{Z}=\int\mathcal{D}\left[\bar{\psi},\psi\right]\exp\left(-S\left[\bar{\psi},\psi\right]\right)$,
where the action is given by $S\left[\bar{\psi},\psi\right]=\int_{0}^{\beta}d\tau\left[\bar{\psi}\partial_{\tau}\psi+H\left(\bar{\psi},\psi\right)\right]$
and $\bar{\psi},\psi$ are independent Grassman variables. 
Natural units, where $\hbar=1,k_{B}=1$, are employed throughout the manuscript; the inverse temperature is then $T^{-1}=\beta$. 
It is advantageous to then apply the Hubbard-Stratonovich (HS) transformation~\cite{Altland_Simons,Taylor_2006, Anderson_2016},
which is an exact statement that eliminates the four-fermion interaction term at the expense of introducing
a functional integral over a bosonic auxiliary field denoted by $\Delta$. 
The HS transformation follows from inserting the resolution of the identity 
$1=\int\mathcal{D}\left[\Delta^{*},\Delta\right]\exp\left(-\int dx\left|\Delta\right|^{2}/g\right)$
and then shifting $\Delta\rightarrow\Delta-g\psi_{\downarrow}\psi_{\uparrow}$.
Here the integration measure is defined such that the functional integral gives unity. 

After introducing the Nambu spinor $\Psi\left(x\right)=\left(\psi_{\uparrow}(x),\bar{\psi}_{\downarrow}\left(x\right)\right)^{T}$,
where $x=\left(\tau,\mathbf{r}\right)$, the complete fermion and HS action can then be written as
\begin{eqnarray}
S_{\mathrm{F}+\mathrm{HS}}\left[\bar{\psi},\psi,\Delta^{*},\Delta\right] & = & \int dxdy\Psi^{\dagger}(x)\left[-\mathcal{G}^{-1}\left(x,y\right)\right]\Psi\left(y\right)\nonumber\\
& + & \int dx\frac{\left|\Delta\left(x\right)\right|^{2}}{g}.
\end{eqnarray}
The inverse Nambu Green's function is defined via Dyson's equation: 
$\mathcal{G}^{-1}\left(x,y\right)=\mathcal{G}_{0}^{-1}\left(x,y\right)-\Sigma\left[\Delta^{*},\Delta\right]\left(x,y\right)$, 
where the bare inverse Nambu Green's function is $\mathcal{G}_{0}^{-1}\left(x,y\right)=\left(-\partial_{\tau}-\xi_{\mathbf{p}}\tau_{z}\right)\delta\left(x-y\right)$, 
and the self energy is 
$\Sigma\left[\Delta^{*},\Delta\right]\left(x,y\right)=-\left(\Delta\left(x\right)\tau_{+}+\Delta^{*}\left(x\right)\tau_{-}\right)\delta\left(x-y\right)$.
Here the single-particle dispersion (in momentum space) is $\xi_{\mathbf{p}}=\mathbf{p}^2/\left(2m\right)-\mu$, 
and $\tau_{x},\tau_{y},\tau_{z}$ are the standard Pauli matrices, with $\tau_{\pm}=\frac{1}{2}\left(\tau_{x}\pm i\tau_{y}\right)$. 
After integrating out the fermions the HS action is obtained~\cite{Altland_Simons,Taylor_2006,Anderson_2016}:
\begin{equation}
S_{\mathrm{HS}}\left[\Delta^{*},\Delta\right] = \int dx\frac{\left|\Delta\left(x\right)\right|^{2}}{g}-\mathrm{Tr\ log}
\left(-\beta\mathcal{G}^{-1}\right).
\end{equation}
The trace operation $\mathrm{Tr}$ denotes a trace over the entire configuration space; 
that is, it is both a trace over Nambu indices (to be denoted by $\mathrm{tr}$) and an integration over spatial coordinates. 

The HS action is an exact expression, however, it is impractical to perform the bosonic functional integral over $S_{\mathrm{HS}}$,
and so suitable approximations must be employed to enable further theoretical tractability. 
The standard assumption is the saddle-point approximation: $\left.\delta S_{\mathrm{HS}}/\delta\Delta^{*}\left(y\right)\right|_{\Delta=\Delta_{\mathrm{mf}}}=0$;
the solution to this equation is the saddle-point (or mean-field) value $\Delta_{\mathrm{mf}}$ of the bosonic auxiliary field. 
Imposing this condition leads to the standard mean-field BCS gap equation:
$\Delta_{\mathrm{mf}}\left(y\right)/g=\mathrm{tr}\left[\mathcal{G}_{\mathrm{mf}}\left(y,y\right)\tau_{-}\right]$, where $\mathcal{G}_{\mathrm{mf}}$ is the mean-field Nambu Green's function. 
The mean-field action at the saddle-point level is thus $S_{\mathrm{mf}}=S_{\mathrm{HS}}\left[\Delta^{*}_{\mathrm{mf}},\Delta_{\mathrm{mf}}\right]$.

\subsection{Gaussian fluctuations}

To go beyond the saddle-point approximation, the bosonic auxiliary field is expanded as follows: 
$\Delta\left(x\right)=\Delta_{\mathrm{mf}}+\eta\left(x\right)$. 
The inverse Nambu-Green's function is then 
$\mathcal{G}^{-1}=\mathcal{G}^{-1}_{\mathrm{mf}}-\Sigma\left[\eta^{*},\eta\right]$, 
where $\Sigma\left[\eta^{*},\eta\right]\left(x,y\right)=-\left(\eta\left(x\right)\tau_{+}+\eta^{*}\left(x\right)\tau_{-}\right)\delta\left(x-y\right)$.
The HS action can then be expanded about the mean-field action, and to quadratic order in $\eta$ the result is:
$S_{\mathrm{HS}}\left[\Delta^{*},\Delta\right]=S_{\mathrm{mf}}\left[\Delta^{*}_{\mathrm{mf}},\Delta_{\mathrm{mf}}\right]
+\int dxdy\left.\frac{1}{2}\eta_{a}\left(x\right)\left(\delta^{2}S_{\mathrm{HS}}/\delta\Delta_{a}\left(x\right)\delta\Delta_{b}\left(y\right)\right)\right|_{\Delta=\Delta_{\mathrm{mf}}}\eta_{b}\left(y\right).$
Here, there is an implicit sum over the indices $a,b$, which denote either $\Delta^{*}, \Delta$. 
Note that, the terms linear in $\eta^{*}$ and $\eta$ vanish due to the saddle-point condition. 
The primary focus of the manuscript is to obtain the fluctuation EM response in the normal state 
and to connect this to the well-known normal-state fluctuation diagrams~\cite{VarlamovBook}.
With this goal in mind, now set $\Delta_{\mathrm{mf}}=0$. 
This then affords many simplifications; for instance, the momentum-space inverse Nambu Green's function becomes
$\mathcal{G}^{-1}\left(k\right)=\mathcal{G}^{-1}_{0}\left(k\right)=i\omega_{n}-\xi_{\mathbf{k}}\tau_{z}$. 
The (free) single-particle Green's function, in momentum space, is defined by $G_{0}\left(k\right)=\left(i\omega_{n}-\xi_{\mathbf{k}}\right)^{-1}$. 

After computing the above second-order derivatives, then taking the trace over the Nambu indices and finally converting to momentum space, 
the following expression is obtained:
$S_{\mathrm{HS}}\left[\Delta^{*},\Delta\right]=S_{0}+\sum_{q}\eta^{*}\left(q\right)\left[-L^{-1}\left(q\right)\right]\eta\left(q\right).$
The first term here is the action of a free (fermionic) system; that is, $S_{0}=-\mathrm{Tr\ log}\left(-\beta\mathcal{G}_{0}^{-1}\right)$.
The momentum-space inverse fluctuation propagator is defined by $L^{-1}\left(q\right)=\Pi\left(q\right)-g^{-1}$, 
where $\Pi\left(q\right)=\sum_{k}G_{0}\left(k\right)G_{0}\left(q-k\right)$ is the pair susceptibility in the Cooper channel. 
The position-space inverse fluctuation propagator is defined by $-L^{-1}\left(x,x^{\prime}\right)=g^{-1}\delta\left(x-x^{\prime}\right)+G_{0}\left(x,x^{\prime}\right)\widetilde{G}_{0}\left(x^{\prime},x\right)$.
Here $G_{0}$ and $\widetilde{G}_{0}$ denote the particle and hole bare propagators~\cite{Altland_Simons}; in momentum space
they are related by $G_{0}\left(k\right)=-\widetilde{G}_{0}\left(-k\right)$. Since the HS action is Gaussian in the (complex) bosonic fields $\eta^{*}, \eta$, the functional integral over these fields can be computed exactly. 
The generating functional is
$\mathcal{Z}=e^{-S_{\mathrm{Eff}}}$ where $S_{\mathrm{Eff}}$ is the effective action, which as a result of the functional integration is then~\cite{Svidzinskii_1971,SvidzinskiiBook}:
\begin{equation}
S_{\mathrm{Eff}}=S_{0}+\mathrm{Tr\ log}\left(-gL^{-1}\right).
\end{equation}
The second term is the Gaussian-fluctuation action, in the absence of a mean-field gap: $S_{\mathrm{Fluc}}=\mathrm{Tr\ log}\left(-gL^{-1}\right)$.
The fluctuation action can also be calculated in the presence of a non-zero mean-field gap, see Ref.~\cite{SvidzinskiiBook} for further details.

\section{Electromagnetic response}
\label{sec:EM_response}
\subsection{Derivation of the fluctuation electromagnetic response kernel}
\label{sec:Fluctuation_diagrams}

The electromagnetic response of the system can be obtained by computing
the action in the presence of an external vector potential, $A^{\mu}\left(x\right)=\left(iA_{0}\left(x\right),\mathbf{A}\left(x\right)\right)$;
here the imaginary unit arises because of the use of imaginary time. 
The resulting fluctuation action is denoted by $S_{\mathrm{Fluc}}\left[A\right]$, and is given by $S_{\mathrm{Fluc}}\left[A\right]=\mathrm{Tr\ log}\left(-gL^{-1}\left[A\right]\right).$
The fluctuation EM response is then determined from the following expression: 
\begin{equation}
K_{\mathrm{Fluc}}^{\mu\nu}\left(x,x^{\prime}\right) = \left.\frac{\delta^{2}S_{\mathrm{Fluc}}\left[A\right]}{\delta A_{\mu}\left(x\right)\delta A_{\nu}\left(x^{\prime}\right)}\right|_{A=0}.
\end{equation}
Note that, in addition to the fluctuation part of the EM response, there is also a mean-field contribution given by $K^{\mu\nu}_{\mathrm{mf}}=\left.\delta^{2}S_{\mathrm{mf}}/\delta A_{\mu}\delta A_{\nu}\right|_{A=0}$. 
The full response is the sum of these two contributions: $K^{\mu\nu}=K_{\mathrm{mf}}^{\mu\nu}+K_{\mathrm{Fluc}}^{\mu\nu}$; 
since the former is well understood~\cite{Arseev_2006,Anderson_2016}, and known to be gauge-invariant: $q_{\mu}K_{\mathrm{mf}}^{\mu\nu}=0$,
the discussion and analysis throughout the remainder of the paper will be devoted to establishing the gauge invariance of the fluctuation part of the response.

To incorporate the vector potential dependence in the fluctuation propagator, the procedure is to perform minimal coupling in the bare Green's function; 
that is, in the particle sector transform $\hat{\mathbf{p}}\rightarrow\hat{\mathbf{p}}-e\mathbf{A}$
and $\partial_{\tau}\rightarrow\partial_{\tau}-ieA_{0}$, while in the hole sector the same transformation is done but with $e\rightarrow-e$.  
Performing the two functional derivatives of the fluctuation action then results in 
\begin{align}
\label{eq:Action_Deriv}
K_{\mathrm{Fluc}}^{\mu\nu}\left(x,x^{\prime}\right)=&
-\int\prod_{i=1}^{4}dy_{i}L\left(y_{1},y_{2}\right)\Lambda^{\mu}\left(y_{2},x,y_{3}\right)\nonumber\\&\quad\times L\left(y_{3},y_{4}\right)\Lambda^{\nu}\left(y_{4},x^{\prime},y_{1}\right)\nonumber\\
&+\int\prod_{i=1}^{2}dy_{i}L\left(y_{1},y_{2}\right)\Gamma^{\mu\nu}\left(y_{2},x,x^{\prime},y_{1}\right).
\end{align}
In the above expression the fluctuation propagator is evaluated at $A_{\mu}=0$; unless otherwise explicitly stated, $L\left(y_{1},y_{2}\right)$ 
denotes $L\left[A=0\right]\left(y_{1},y_{2}\right)$, with a similar relation for the Green's functions.
The three-point and four-point vertices are respectively defined by 
\begin{eqnarray}
\Lambda^{\mu}\left(y_{1},x,y_{2}\right)&=&\left.\frac{\delta L^{-1}\left[A\right]\left(y_{1},y_{2}\right)}{\delta A_{\mu}\left(x\right)}\right|_{A=0},\label{eq:AL_vertex}\\
\Gamma^{\mu\nu}\left(y_{1},x,x^{\prime},y_{2}\right)&=&\left.\frac{\delta^{2}L^{-1}\left[A\right]\left(y_{1},y_{2}\right)}{\delta A_{\mu}\left(x\right)\delta A_{\nu}\left(x^{\prime}\right)}\right|_{A=0}.
\end{eqnarray}
Note that these vertices are independent of the external vector potential. 
In Eq.~(\ref{eq:Action_Deriv}) the first term is the AL diagram, while the second term contains the MT and DOS diagrams, 
plus an additional Gaussian-level diamagnetic contribution, as will be shown forthwith. 

The second term above can be written schematically as:
\begin{eqnarray}
\label{eq:Double_Deriv}
&&-\left.\frac{\delta^{2}L^{-1}\left[A\right]}{\delta A_{\mu}\delta A_{\nu}}\right|_{A=0} = 2\biggl\{G_{0}\frac{\delta G_{0}^{-1}}{\delta A_{\nu}}G_{0}\widetilde{G}_{0}\frac{\delta \widetilde{G}_{0}^{-1}}{\delta A_{\mu}}\widetilde{G}_{0}\nonumber \\
&+&G_{0}\frac{\delta G_{0}^{-1}}{\delta A_{\mu}}G_{0}\frac{\delta G_{0}^{-1}}{\delta A_{\nu}}G_{0}\widetilde{G}_{0}+G_{0}\frac{\delta G_{0}^{-1}}{\delta A_{\nu}}G_{0}\frac{\delta G_{0}^{-1}}{\delta A_{\mu}}G_{0}\widetilde{G}_{0}\nonumber \\
&-&\left.G_{0}\frac{\delta^{2}G_{0}^{-1}}{\delta A_{\mu}\delta A_{\nu}}G_{0}\widetilde{G}_{0}\biggr\}\right|_{A=0}.
\end{eqnarray}
The prefactor of two here arises due to the particle-hole symmetry present in the Nambu formalism.
It is implicit in this expression that there is a ``matrix multiplication", with the ``multiplication"  between neighboring terms corresponding to an integration over coordinates. 
In Eq.~(\ref{eq:Double_Deriv}), the first term is the MT diagram, the second and third terms are the two DOS diagrams, and the fourth term is an additional contribution. 
This latter term represents a (momentum-independent) Gaussian-level diamagnetic contribution to the EM response.
However, it contributes to only the current-current part of the EM response.
Note that, here it has been demonstrated that the MT, DOS, and the Gaussian-level diamagnetic diagrams all arise from the second term in Eq.~(\ref{eq:Action_Deriv}). 
This result is not easily apparent in the conventional derivation of the MT and DOS diagrams~\cite{VarlamovBook}, 
and so this shows an advantage in utilizing the functional-integral approach. 
It concretely shows that all diagrams, and in particular MT, DOS, plus the Gaussian-level diamagnetic term, are on an equal footing as respects their origin within the Gaussian-fluctuation theory.

The (particle) bare vertex appearing above is defined by
$\left.\delta G_{0}^{-1}\left[A\right]\left(y_{1},y_{2}\right)/\delta A_{\mu}\left(x\right)\right|_{A=0}\equiv e\gamma^{\mu}\left(y_{1},x,y_{2}\right)$.
Its Fourier transform is then 
$\gamma^{\mu}\left(y_{1},x,y_{2}\right)=\sum_{k_{1},k_{2}}e^{ik_{1}\left(y_{1}-y_{2}\right)+ik_{2}\left(y_{1}-x\right)}\gamma^{\mu}\left(k_{1}+k_{2},k_{1}\right)$, 
where the momentum-space bare vertex is~\cite{Schrieffer} $\gamma^{\mu}\left(k_{1}+k_{2},k_{1}\right)=\left(1,\frac{1}{m}\left[\mathbf{k}_{1}+\mathbf{k}_{2}/2\right]\right)$
\footnote{A factor of $e$ has been pulled out in the definition of the vertices to facilitate later comparison with the standard expressions for the fluctuation diagrams.}. 
The hole bare vertex is related to the particle bare vertex by $\widetilde{\gamma}^{\mu}\left(k_{1}+k_{2},k_{1}\right)=-\gamma^\mu\left(-k_{1},-k_{1}-k_{2}\right)$. 
Converting the EM response to momentum space gives
\begin{equation}\label{eq:Fluc_EM}
K_{\mathrm{Fluc}}^{\mu\nu}\left(x,x^{\prime}\right) = \sum_{q}e^{-iq\left(x-x^{\prime}\right)}K_{\mathrm{Fluc}}^{\mu\nu}\left(q\right),
\end{equation}
where the fluctuation EM response kernel is 
\begin{align}
K_{\mathrm{Fluc}}^{\mu\nu}\left(q\right) &= -\sum_{p}L\left(p+q\right)\Lambda^{\mu}\left(p+q,p\right)L\left(p\right)\Lambda^{\nu}\left(p,p+q\right)\nonumber\\
&\quad + \sum_{p}L\left(p\right)\Gamma^{\mu\nu}\left(p,q\right).
\end{align}
Here $q^{\mu}=\left(i\Omega_{m},\mathbf{q}\right)$, $p^{\mu}=\left(i\epsilon_{m},\mathbf{p}\right)$, where $\epsilon_{m}$ and $\Omega_{m}$ are bosonic Matsubara frequencies, 
with $i\Omega_{m}=\Omega+i0^{+}$ the analytic continuation of the latter. The four-vector summation is defined by $\sum_{p}\equiv (\beta V)^{-1}\sum_{i\epsilon_{m}}\sum_{\mathbf{p}}$,
where $V$ is the volume and $V^{-1}\sum_{\mathbf{p}}\equiv\int d^{d}\mathbf{p}/\left(2\pi\right)^{d}$. 
Note that $K^{\mu\nu}\left(q\right)=K^{\nu\mu}\left(-q\right)$, and so there are different, but equivalent, ways of writing the EM response. 
As shown in the appendices~(\ref{sec:App_AL}-\ref{sec:App_GD}), after Fourier transforming the position-space expressions the fluctuation response in momentum space can be written as: 
\begin{eqnarray}
\label{eq:Fluctuation_response}
K_{\mathrm{Fluc}}^{\mu\nu}\left(q\right) & = & \mathrm{AL}^{\mu\nu}\left(q\right)+\mathrm{MT}^{\mu\nu}\left(q\right)\nonumber\\
& & + \mathrm{DOS}^{\mu\nu}\left(q\right)+\delta K^{\mu\nu}_{\mathrm{Fluc}}\left(q\right),
\end{eqnarray}
where the explicit expressions for these diagrams are 
\begin{widetext}
\begin{eqnarray}
\mathrm{AL}^{\mu\nu}\left(i\Omega_{m},\mathbf{q}\right) & = & -4e^{2}\sum_{k,l,p}G_{0}\left(l+q\right)\gamma^{\mu}\left(l+q,l\right)G_{0}\left(l\right)G_{0}\left(p-l\right)L\left(p+q\right)L\left(p\right)\nonumber \\
 &  & \times G_{0}\left(p-k\right)G_{0}\left(k\right)\gamma^{\nu}\left(k,k+q\right)G_{0}\left(k+q\right),\label{eq:AL_diagram}\\
\mathrm{MT}^{\mu\nu}\left(i\Omega_{m},\mathbf{q}\right) & = & 2e^{2}\sum_{k,p}G_{0}\left(p-k\right)\gamma^{\mu}\left(p-k,p-k-q\right)G_{0}\left(p-k-q\right)L\left(p\right)G_{0}\left(k\right)\gamma^{\nu}\left(k,k+q\right)G_{0}\left(k+q\right),\label{eq:MT_diagram}\\
\mathrm{DOS}{}^{\mu\nu}\left(i\Omega_{m},\mathbf{q}\right) & = & 2e^{2}\sum_{k,p}G_{0}\left(k+q\right)\gamma^{\mu}\left(k+q,k\right)G_{0}\left(k\right)\gamma^{\nu}\left(k,k+q\right)G_{0}\left(k+q\right)L\left(p\right)G_{0}\left(p-k-q\right)\nonumber\\
 & & + 2e^{2}\sum_{k,p}G_{0}\left(k+q\right)\gamma^{\mu}\left(k+q,k\right)G_{0}\left(k\right)L\left(p\right)G_{0}\left(p-k\right)G_{0}\left(k\right)\gamma^{\nu}\left(k,k+q\right),\label{eq:DOS_diagrams}\\
\delta K^{\mu\nu}_{\mathrm{Fluc}}\left(i\Omega_{m},\mathbf{q}\right) & = & \frac{2e^2}{m}\delta^{\mu i}\delta^{\nu i}\sum_{k,p}L\left(p\right)G_{0}\left(p-k\right)G_{0}^{2}\left(k\right).\label{eq:Gauss_diamag_diagram}
\end{eqnarray}
\end{widetext}

Here  $k^{\mu}=\left(i\omega_{n},\mathbf{k}\right),l^{\mu}=\left(i\varphi_{n},\mathbf{l}\right)$, where $\omega_{n}$ and $\varphi_{n}$ are fermionic Matsubara frequencies. 
The first three terms are the familiar expressions~\cite{VarlamovBook} for the AL, MT, and DOS diagrams. 
What is of interest here is the additional term given by $\delta K^{\mu\nu}_{\mathrm{Fluc}}$, which is not present in the papers of Aslamazov and Larkin~\cite{Aslamazov_1968, Aslamazov_1975}. 
A similar diagram can be found in Fig.~(11) of Ref.~\cite{SachdevHartnoll_2011} in the context of cross-correlations between hot and cold fermion currents, and it was noted there that this three-point function is in general finite. 
This term is a purely real three-point function. It is independent of external momentum, and so as a consequence it does not contribute to finite-momentum or finite-frequency transport.
Thus the results of Aslamazov and Larkin for the finite-frequency conductivity and diamagnetic susceptibility are unaltered. The electrical conductivity is discussed in further detail in subsection~(\ref{sec:Conductivity}).
In Appendix~(\ref{sec:Diamag_diagrams}) it is proved that the Gaussian-level diamagnetic diagram can be expressed simply in terms of the partial derivative of the fluctuation action with respect to the fermionic chemical potential as 
$\delta K^{\mu\nu}_{\mathrm{Fluc}}\left(i\Omega_{m},\mathbf{q}\right)=-\delta^{\mu i}\delta^{\nu i}\frac{e^{2}}{m}\frac{1}{\beta V}\frac{\partial S_{\mathrm{Fluc}}}{\partial\mu}$.
The next two subsections show that this diagram is important for establishing (i) gauge invariance, and (ii) the absence of the normal-state Meissner effect.

The Feynman diagrams for the complete fluctuation EM response kernel are shown in Fig.~(\ref{fig:Fluctuation_diagrams}). Note that, these diagrams are not an expansion in loops, or equivalently in terms of the number of fluctuation propagators. This was in fact the reason why Maki missed the AL diagram in his paper~\cite{Maki_1968}, as he was focused solely on the single-fluctuation-propagator diagrams (b)-(d). The diagrams in Fig.~(\ref{fig:Fluctuation_diagrams}) are the complete set of (ultraclean) Gaussian-level fluctuation diagrams. It is the action which has been computed at the Gaussian level, or equivalently, in the diagrammatic language, the free energy has been calculated with only a single fluctuation propagator~\cite{AGDBook,VarlamovBook}. 

\begin{figure*}[t]
\centering\includegraphics[width=12cm,height=8cm,clip]{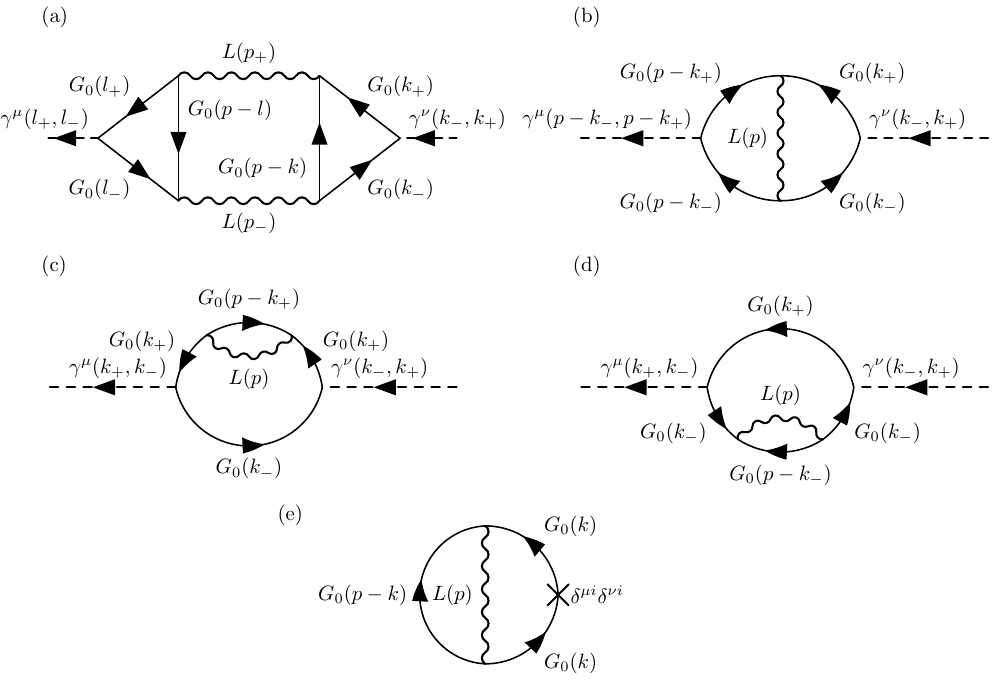}
\caption{The complete set of Gaussian-level fluctuation diagrams in the normal state. In order, these diagrams correspond to (a) Aslamazov-Larkin, (b) Maki-Thompson, (c) and (d) Density of States, and (e) Gaussian-level diamagnetic. The solid lines are bare electron propagators and the wiggly lines are fluctuation propagators.
Here $k_{+}=k+q,k_{-}=k$, etc.}
\label{fig:Fluctuation_diagrams}
\end{figure*}

For completeness, it is useful to mention how the diagram in Eq.~(\ref{eq:Gauss_diamag_diagram}) arises within the standard diagrammatic technique. 
A derivation of the AL, MT, and DOS diagrams from the fluctuation free energy is presented on page 117 of Ref. ~\cite{VarlamovBook}. However, the diagram derived in Eq.~(\ref{eq:Gauss_diamag_diagram}) is absent. Nevertheless, it can be obtained by carefully following that references methodology. The fluctuation free energy diagram (see Fig.~6.2(a) in Ref.~\cite{VarlamovBook}) consists of a closed loop of two bare electron propagators connected by a single fluctuation propagator. A single functional derivative of this diagram, with respect to an external source, results in attaching a bare vertex in all possible positions, and after summing the resultant contributions a single bare vertex is attached to one of the bare electron propagators (see Fig.~6.2(b) in Ref.~\cite{VarlamovBook}). Performing a second functional derivative of this diagram results in the complete fluctuation response; the derivatives of the fluctuation propagator and the electron propagators give rise to the AL, MT, and DOS diagrams (see Fig.~6.2(c) in Ref.~\cite{VarlamovBook}). However, what was neglected in that reference was to differentiate the external vertex as well. The derivative of the external vertex, with respect to the external source, amounts to detaching the vertex from the diagram~\cite{Ryder} and inserting a Kronecker delta function, the result of which exactly produces the diagram in Eq.~(\ref{eq:Gauss_diamag_diagram}). The Gaussian-level diamagnetic diagram, absent from Refs.~\cite{Aslamazov_1968,Aslamazov_1975,Aslamazov_1980,Aslamazov2_1980,Svidzinskii_1971,SvidzinskiiBook,VarlamovBook,Varlamov_2018}, can thus be derived with the functional-integral approach, as showed in this section, and also with the diagrammatic technique, as just discussed.

\subsection{Proving gauge invariance}

In this subsection it is proved that the normal-state fluctuation response is gauge-invariant; this is mathematically expressed by the statement that $q_{\mu}K_{\mathrm{Fluc}}^{\mu\nu}\left(q\right)=0$. 
It will be shown that this is only true provided the Gaussian-level diamagnetic contribution is incorporated. 
Note that, in Ref.~\cite{Anderson_2016} the full Gaussian-level response was derived, with the inclusion of the anomalous Green's function and the collective modes. 
There it was shown, in a more abstract manner, that provided the dependence of the order-parameter on the external vector potential is incorporated, gauge invariance is maintained. 
Here only the normal-state response is considered, and the gauge invariance of the theory will be demonstrated by explicitly calculating the four-vector contractions of all the fluctuation diagrams. 
To perform the requisite contractions, use is made of the bare Ward-Takahashi identity~\cite{Ryder}: 
$q_{\mu}\gamma^{\mu}\left(k+q,k\right)=G_{0}^{-1}\left(k+q\right)-G_{0}^{-1}\left(k\right)$. 
Note that, using the definition of the bare vertex, $\gamma^{\mu}\left(k+q,k\right)=\left(1,\frac{1}{m}\left[\mathbf{k}+\mathbf{q}/2\right]\right)$, this identity is easily proved.  

Using the bare Ward-Takahashi identity, it follows that the contractions of the MT, DOS, and AL diagrams are 
\begin{widetext}
\begin{align}
q_{\mu}\mathrm{MT}^{\mu\nu}\left(i\Omega_{m},\mathbf{q}\right)  & =  2e^{2}\sum_{k,p}\left[G_{0}\left(p-k-q\right)-G_{0}\left(p-k\right)\right]L\left(p\right)G_{0}\left(k\right)\gamma^{\nu}\left(k,k+q\right)G_{0}\left(k+q\right).\\
q_{\mu}\mathrm{DOS}{}^{\mu\nu}\left(i\Omega_{m},\mathbf{q}\right) & =  2e^{2}\sum_{k,p}G_{0}\left(k+q\right)G_{0}\left(p-k-q\right)L\left(p\right)\left[G_{0}\left(k\right)-G_{0}\left(k+q\right)\right]\gamma^{\nu}\left(k,k+q\right)\nonumber \\
 & \quad + 2e^{2}\sum_{k,p}\left[G_{0}\left(k\right)-G_{0}\left(k+q\right)\right]G_{0}\left(p-k\right)L\left(p\right)G_{0}\left(k\right)\gamma^{\nu}\left(k,k+q\right).\\
q_{\mu}\mathrm{AL}^{\mu\nu}\left(i\Omega_{m},\mathbf{q}\right) & = -4e^{2}\sum_{k,l,p}\left[G_{0}\left(l\right)-G_{0}\left(l+q\right)\right]G_{0}\left(p-l\right)L\left(p+q\right)L\left(p\right)G_{0}\left(p-k\right)G_{0}\left(k\right)
\gamma^{\nu}\left(k,k+q\right)G_{0}\left(k+q\right),\nonumber \\
 & = -4e^{2}\sum_{k,p}\left[L\left(p+q\right)-L\left(p\right)\right]G_{0}\left(p-k\right)G_{0}\left(k\right)\gamma^{\nu}\left(k,k+q\right)G_{0}\left(k+q\right),\nonumber \\
 & = -4e^{2}\sum_{k,p}\left[G_{0}\left(p-k-q\right)-G_{0}\left(p-k\right)\right]L\left(p\right)G_{0}\left(k\right)\gamma^{\nu}\left(k,k+q\right)G_{0}\left(k+q\right).
\end{align}
Here we have used the definition of the inverse fluctuation propagator, $L^{-1}\left(p\right)+g^{-1}=\Pi\left(p\right)=\sum_{l}G_{0}\left(l\right)G_{0}\left(p-l\right)$, to perform the fermionic Matsubara frequency summation over $l$. 
Combining the above results, and simplifying, then gives
\begin{eqnarray}
\label{eq:Contractions}
q_{\mu}\left[\mathrm{AL}^{\mu\nu}\left(q\right)+\mathrm{MT}^{\mu\nu}\left(q\right)+\mathrm{DOS}^{\mu\nu}\left(q\right)\right] & = & 2e^{2}\sum_{k,p}G_{0}\left(k\right)L\left(p\right)G_{0}\left(p-k\right)\left[\gamma^{\nu}\left(k,k+q\right)-\gamma^{\nu}\left(k-q,k\right)\right]G_{0}\left(k\right),\nonumber \\
 & = & 2e^{2}\frac{\mathbf{q}}{m}\sum_{k,p}L\left(p\right)G_{0}\left(p-k\right)G_{0}^{2}\left(k\right),\nonumber\\
 & = & - q_{\mu}\delta K^{\mu\nu}_{\mathrm{Fluc}}\left(q\right),
\end{eqnarray}
\end{widetext}
where in the last step the contraction of Eq.~(\ref{eq:Gauss_diamag_diagram}) is easily computed. 
As a result, the contraction of the fluctuation EM response kernel is 
\begin{equation}
q_{\mu}K_{\mathrm{Fluc}}^{\mu\nu}\left(q\right)  = 0.
\end{equation}
Therefore the fluctuation EM response is gauge invariant. It is important to note that, without incorporating the Gaussian-level diamagnetic diagram, the fluctuation response would not be gauge-invariant,
as shown in the second line of Eq.~(\ref{eq:Contractions}). The next subsection investigates the Meissner response of the fluctuation EM response kernel.

\subsection{Absence of the Meissner effect in the normal state}
\label{sec:Meissner}

The superfluid density of a system is a static, zero momentum response, and in the normal state of a superconductor any
gauge-invariant calculation must produce zero Meissner response~\cite{Schafroth}. 
In this subsection it will be proved that the exact normal-state fluctuation response does not exhibit the Meissner effect:
$K^{ij}\left(\Omega=0,\mathbf{q}\rightarrow0\right)=K_{\mathrm{Fluc}}^{ij}\left(\Omega=0,\mathbf{q}\rightarrow0\right)=0$.
As will be shown, this result is only true with the incorporation of the Gaussian-level diamagnetic contribution. 
This reiterates the point that the additional diagram derived in 
Sec.~(\ref{sec:Fluctuation_diagrams}) represents an important
contribution to the full EM response and in general it cannot be neglected.

By using the identity~\cite{Ryder} $G_{0}\left(k\right)\gamma^{i}\left(k,k\right)G_{0}\left(k\right)=\partial G_{0}\left(k\right)/\partial k^{i}$,
along with the relation $\Pi\left(p\right)=\sum_{k}G_{0}\left(k\right)G_{0}\left(p-k\right)=L^{-1}\left(p\right)+g^{-1}$,
the fermionic Matsubara frequency summations appearing in the fluctuation diagrams can be performed. 
The final results for the static fluctuation diagrams, in the zero-momentum limit, are 
\begin{eqnarray}
\mathrm{AL}^{ij}\left(\Omega=0,\mathbf{q}\rightarrow0\right) & = & -4e^{2}\sum_{p}L\left(p\right)\frac{\partial^{2}\Pi\left(p\right)}{\partial p^{i}\partial p^{j}},\nonumber\\
\mathrm{MT}^{ij}\left(\Omega=0,\mathbf{q}\rightarrow0\right) & = & 2e^{2}\sum_{p}L\left(p\right)\frac{\partial^{2}\Pi\left(p\right)}{\partial p^{i}\partial p^{j}},\nonumber\\
\mathrm{DOS}^{ij}\left(\Omega=0,\mathbf{q}\rightarrow0\right) & = & 2e^{2}\sum_{p}L\left(p\right)\frac{\partial^{2}\Pi\left(p\right)}{\partial p^{i}\partial p^{j}}\nonumber\\
 & & - \delta K^{ij}_{\mathrm{Fluc}}.\end{eqnarray}
Adding these terms together, along with the contribution from Eq.~(\ref{eq:Gauss_diamag_diagram}), 
it follows that the zero-momentum limit of the static fluctuation EM response kernel is thus
\begin{equation}
K_{\mathrm{Fluc}}^{ij}\left(\Omega=0,\mathbf{q}\rightarrow0\right) = 0.
\end{equation}
This proves that the normal-state fluctuation EM response does not exhibit the Meissner effect, as required. 
It is important to emphasize, that it is only with the inclusion of the diagram in Eq.~(\ref{eq:Gauss_diamag_diagram}) that there is no anomalous diamagnetism; this corrects
Refs.~\cite{Aslamazov_1980, Aslamazov2_1980,VarlamovBook}, which asserted that the AL, MT, and DOS diagrams were sufficient.

\section{Electrical conductivity}
\label{sec:Conductivity}
The fluctuation contribution to the electrical conductivity of an ultraclean layered 2D superconductor, with no impurities or disorder and constant bare vertices, was computed in Ref.~\cite{Varlamov_1991}; it was found that (when averaging over the Fermi surface) the conductivities of the MT and DOS contributions cancel one another, a result which was also obtained in Ref.~\cite{SachdevHartnoll_2011} in a different context. A subtle issue arose in Ref.~\cite{Varlamov_1991}, where it was proved that the triangle vertex in the ultraclean AL diagram is non-zero only when the external Matsubara frequency is zero, and for all other non-zero frequencies the triangle vertex vanishes. As a consequence, the validity of analytic continuation of such a function was brought into question. Later on this issue was resolved in Ref.~\cite{Larkin_1995}, where it was shown that the AL diagram is a non-local function, and in the ultraclean limit the non-local AL conductivity vanishes at all non-zero frequencies.

The conductivities of systems with impurities have also been addressed~\cite{Larkin_1995, Axnas_1999, Varlamov_2000,Galitski_2001,VarlamovBook,Finkelstein_2012,Fiegelman_2018,Skvortsov_2018}, 
and there has been some debate about whether the clean limit of an impure superconductor reproduces the ultraclean results; that is, it is debated whether or not
the order of the limits $\Omega\rightarrow0,\tau\rightarrow\infty$ and $\tau\rightarrow\infty,\Omega\rightarrow0$ commute with one another in the response functions. 
Here $\tau$ is a lifetime for particle scattering. For the non-local AL conductivity, it was noted~\cite{Larkin_1995} that the $\tau$ dependence is of the form $\sim\left(1-i\Omega\tau\right)^{-2}$, 
and it was physically argued that the appearance of a second-order diffusive pole (in contrast to the first-order pole familiar in the Drude theory of a free system) is because  fluctuation Cooper pairs do not directly interact with an applied field; rather, the applied field interacts with electrons, which in turn induce a fluctuation response with an additional time retardation $\tau$. 

In the present paper only the ultraclean case has been considered hitherto, where impurities and disorder are absent from the outset.
In this section it will be proved that the electrical conductivities of the ultraclean MT,  DOS , and AL diagrams exactly sum to zero. 
Thus, the ultraclean fluctuation conductivity is due solely to the Gaussian-level diamagnetic diagram obtained in Eq.~(\ref{eq:Gauss_diamag_diagram}). 

To begin, recall that the bare Ward-Takahashi identity is  $q_{\mu}\gamma^{\mu}\left(k+q,k\right)=G_{0}^{-1}\left(k+q\right)-G_{0}^{-1}\left(k\right).$
Setting the momentum ${\bf q}=0$ and then rearranging gives:
\begin{align}
&G_{0}\left(i\omega_{n}+i\Omega_{m},{\bf k}\right)G_{0}\left(i\omega_{n},{\bf k}\right)
=\nonumber\\&-\frac{1}{i\Omega_{m}}\left[G_{0}\left(i\omega_{n}+i\Omega_{m},{\bf k}\right)-G_{0}\left(i\omega_{n},{\bf k}\right)\right].\label{eq:Identity1}
\end{align}
This identity will be used to prove the ultraclean MT, DOS, and AL fluctuation diagrams cancel one another. Consider first the MT diagram given in Eq.~(\ref{eq:MT_diagram}). 
Using the result in Eq.(\ref{eq:Identity1}), this becomes 
\begin{widetext}
\begin{eqnarray}
\label{eq:MTCond}
\text{MT}^{xx}\left(i\Omega_{m},\mathbf{0}\right) &  =  & 2e^{2}\left(\frac{1}{i\Omega_{m}}\right)^{2}\sum_{k,p}\gamma^{x}\left(p-k,p-k\right)\gamma^{x}\left(k,k\right)G_{0}\left(p-k\right)G_{0}\left(k\right)\nonumber \\
 &  & \times\left[L\left(i\epsilon_{m}+i\Omega_{m},{\bf p}\right)+L\left(i\epsilon_{m}-i\Omega_{m},{\bf p}\right)-2L\left(i\epsilon_{m},{\bf p}\right)\right].
\end{eqnarray}

The DOS diagrams are given in Eq.~(\ref{eq:DOS_diagrams}), and by using the result in Eq.(\ref{eq:Identity1}) twice they become
\begin{eqnarray}
\label{eq:DOSCond}
\text{DOS}^{xx}\left(i\Omega_{m},\mathbf{0}\right) & = & 2e^{2}\left(\frac{1}{i\Omega_{m}}\right)^{2}\sum_{k,p}\gamma^{x}\left(k,k\right)\gamma^{x}\left(k,k\right)G_{0}\left(p-k\right)G_{0}\left(k\right)\nonumber \\
 &  & \times\left[L\left(i\epsilon_{m}+i\Omega_{m},{\bf p}\right)+L\left(i\epsilon_{m}-i\Omega_{m},{\bf p}\right)-2L\left(i\epsilon_{m},{\bf p}\right)\right].
\end{eqnarray}

Finally, consider the AL diagram given in Eq.~(\ref{eq:AL_diagram}). The triangle vertices in the AL diagram are [see Eq.~(\ref{eq:Lambda_vertex})]
\begin{eqnarray}
\Lambda^{x}\left(i\epsilon_{m}+i\Omega_{m},\mathbf{p};i\epsilon_{m},\mathbf{p}\right) & = & -2e\sum_{k}G_{0}\left(i\omega_{n}+i\Omega_{m},\mathbf{k}\right)\gamma^{x}\left(k,k\right)G_{0}\left(k\right)G_{0}\left(p-k\right), \nonumber\\ 
& = & \frac{2e}{i\Omega_{m}}\sum_{k}\left[G_{0}\left(i\omega_{n}+i\Omega_{m},{\bf k}\right)-G_{0}\left(i\omega_{n},{\bf k}\right)\right]\gamma^{x}\left(k,k\right)G_{0}\left(p-k\right).
\end{eqnarray}
Here $\mathbf{q}=0$. The sums appearing above can be computed as follows. 
\begin{eqnarray}
\sum_{k}G_{0}\left(i\omega_{n}+i\Omega_{m},{\bf k}\right)\gamma^{x}\left(k,k\right)G_{0}\left(p-k\right)
 & = & \sum_{k}G_{0}\left(k\right)\gamma^{x}\left(-k+p,-k+p\right)G_{0}\left(i\epsilon_{m}+i\Omega_{m}-i\omega_{n},{\bf p}-{\bf k}\right),\nonumber \\
 & = & -\sum_{k}G_{0}\left(i\omega_{n}+i\Omega_{m},{\bf k}\right)\gamma^{x}\left(k,k\right)G_{0}\left(p-k\right)\nonumber \\
  &  & + \gamma^{x}\left(p,p\right)\sum_{k}G_{0}\left(k\right)G_{0}\left(i\epsilon_{m}+i\Omega_{m}-i\omega_{n},{\bf p}-{\bf k}\right), \nonumber \\
 & = & \frac{1}{2}\gamma^{x}\left(p,p\right)\sum_{k}G_{0}\left(k\right)G_{0}\left(i\epsilon_{m}+i\Omega_{m}-i\omega_{n},{\bf p}-{\bf k}\right),\nonumber \\
 & = & \frac{1}{2}\gamma^{x}\left(p,p\right)\Pi\left(i\epsilon_{m}+i\Omega_{m},\mathbf{p}\right).
\end{eqnarray}
Similarly, 
\begin{equation}
\label{eq:Identity2}
\sum_{k}G_{0}\left(i\omega_{n},{\bf k}\right)\gamma^{x}\left(k,k\right)G_{0}\left(p-k\right) = \frac{1}{2}\gamma^{x}\left(p,p\right)\Pi\left(p\right).
\end{equation}
The triangle vertices in the AL diagram are thus 
\begin{equation}
\Lambda^{x}\left(p+q,p\right)=\frac{e}{i\Omega_{m}}\gamma^{x}\left(p,p\right)\left[\Pi\left(p+q\right)-\Pi\left(p\right)\right].
\end{equation}
Therefore, the AL diagram is given by
\begin{eqnarray}
\text{AL}^{xx}\left(i\Omega_{m},\mathbf{0}\right) & = & \frac{2e^{2}}{i\Omega_{m}}\sum_{k,p}G_{0}\left(k+q\right)\gamma^{x}\left(k,k\right)G_{0}\left(k\right)G_{0}\left(p-k\right)L\left(p+q\right)L\left(p\right)\gamma^{x}\left(p,p\right)\left[\Pi\left(p+q\right)-\Pi\left(p\right)\right],\nonumber \\
 & = & -2e^{2}\left(\frac{1}{i\Omega_{m}}\right)^{2}\sum_{k,p}\gamma^{x}\left(k,k\right)\left[\gamma^{x}\left(p-k,p-k\right)+\gamma^{x}\left(k,k\right)\right]\left[G_{0}\left(i\omega_{n}+i\Omega_{m},{\bf k}\right)-G_{0}\left(i\omega_{n},{\bf k}\right)\right]\nonumber \\
 &  & \times G_{0}\left(p-k\right)L\left(p+q\right)L\left(p\right)\left[\Pi\left(p+q\right)-\Pi\left(p\right)\right].
\end{eqnarray}
Simplifying this using $L^{-1}\left(p\right)=\Pi\left(p\right)-g^{-1}$ then gives 
\begin{eqnarray}
\label{eq:ALCond}
\text{AL}^{xx}\left(i\Omega_{m},\mathbf{0}\right) & = & -2e^{2}\left(\frac{1}{i\Omega_{m}}\right)^{2}\sum_{k,p}\gamma^{x}\left(k,k\right)\left[\gamma^{x}\left(p-k,p-k\right)+\gamma^{x}\left(k,k\right)\right]\nonumber \\
 &  & \times G_{0}\left(p-k\right)\left[G_{0}\left(i\omega_{n}+i\Omega_{m},{\bf k}\right)-G_{0}\left(i\omega_{n},{\bf k}\right)\right]\left[L\left(p\right)-L\left(p+q\right)\right], \nonumber\\
 & = & -2e^{2}\left(\frac{1}{i\Omega_{m}}\right)^{2}\sum_{k,p}\gamma^{x}\left(k,k\right)\gamma^{x}\left(p-k,p-k\right)G_{0}\left(p-k\right)G_{0}\left(k\right)\nonumber \\
 &  & \times\left[L\left(i\epsilon_{m}+i\Omega_{m},{\bf p}\right)+L\left(i\epsilon_{m}-i\Omega_{m},{\bf p}\right)-2L\left(i\epsilon_{m},{\bf p}\right)\right]\nonumber \\
 &  & -2e^{2}\left(\frac{1}{i\Omega_{m}}\right)^{2}\sum_{k,p}\gamma^{x}\left(k,k\right)\gamma^{x}\left(k,k\right)G_{0}\left(p-k\right)G_{0}\left(k\right)\nonumber \\
 &  & \times\left[L\left(i\epsilon_{m}+i\Omega_{m},{\bf p}\right)+L\left(i\epsilon_{m}-i\Omega_{m},{\bf p}\right)-2L\left(i\epsilon_{m},{\bf p}\right)\right].
\end{eqnarray}
Thus, by combining Eqs.~(\ref{eq:MTCond}, \ref{eq:DOSCond}, \ref{eq:ALCond}), the sum of the MT, DOS, and AL diagrams then vanishes
\begin{equation}
\text{MT}^{xx}\left(i\Omega_{m},\mathbf{0}\right)+\text{DOS}^{xx}\left(i\Omega_{m},\mathbf{0}\right)+\text{AL}^{xx}\left(i\Omega_{m},\mathbf{0}\right) = 0.
\end{equation}
\end{widetext}
This proof of the cancellation of the MT, DOS, and AL contributions made no assumptions about Fermi-surface averaging. 
 
The Kubo formula for the frequency-dependent electrical conductivity is~\cite{Altland_Simons}:
\begin{equation}
\sigma^{ij}\left(\Omega\right)=\underset{\mathbf{q}\rightarrow0}{\mathrm{lim}}i\frac{K^{ij}\left(\Omega,\mathbf{q}\right)}{\Omega+i0^{+}}.
\end{equation}
It is crucial to first take the limit $\mathbf{q}\rightarrow0$; as the previous section shows, first setting the external frequency to zero and then taking $\mathbf{q}\rightarrow0$ corresponds to the Meissner response. 
The only contribution to the fluctuation electrical conductivity is due to Eq.~(\ref{eq:Gauss_diamag_diagram}). 
Since $\delta K^{xx}_{\mathrm{Fluc}}$ has no momentum dependence, its contribution to the longitudinal electrical conductivity is thus
\begin{equation}
\sigma^{xx}_{\mathrm{Fluc}}\left(\Omega\right)=\frac{i\delta K^{xx}_{\mathrm{Fluc}}}{\Omega+i0^{+}}.
\end{equation}
The real part of this expression is $\Re\sigma^{xx}_{\mathrm{Fluc}}\left(\Omega\right)=\pi\delta K^{xx}_{\mathrm{Fluc}}\delta\left(\Omega\right)$. 
This result is analogous to the case of a free-particle system, with no impurities or disorder. In that case the real part of the total conductivity is given by 
$\Re\sigma^{xx}_{\mathrm{Free}}\left(\Omega\right)=n\left(\pi e^2/m\right)\delta\left(\Omega\right)$~\cite{MahanBook}; in a free system, with no mechanism for particle scattering, the conductivity vanishes at non-zero frequencies, 
and to ensure that the conductivity sum rule~\cite{MahanBook} is satisfied all the Drude weight is localized at zero frequency. If impurities are present, then the delta function result is smeared out into a Lorentzian distribution.

As discussed in Sec.~(\ref{sec:Fluctuation_diagrams}), the fluctuation diagram $\delta K^{xx}_{\mathrm{Fluc}}$ can be written as $\delta K^{xx}_{\mathrm{Fluc}}=n_{\mathrm{Fluc}}e^2/m$. 
Similarly, the free EM response, arising from the action $S_{0}$, has a diamagnetic term given by $\delta K^{xx}_{0}=n_{0}e^2/m$, where $n_{0}$ is the free-electron particle density; the total electrical conductivity from the free and fluctuation diamagnetic contributions is thus $\left(n_{0}+n_{\mathrm{Fluc}}\right)\left(\pi e^2/m\right)\delta\left(\Omega\right)$. 
Thus the conductivity sum rule for the ultraclean fluctuation theory is satisfied:
\begin{align}
\int^{\infty}_{-\infty}\frac{d\Omega}{\pi}\Re\sigma^{xx}(\Omega)&=\int^{\infty}_{-\infty}\frac{d\Omega}{\pi}(n_{0}+n_{\mathrm{Fluc}})(\pi e^2/m)\delta(\Omega),\nonumber\\
&=(n_{0}+n_{\mathrm{Fluc}})(e^{2}/m)\equiv ne^{2}/m.
\end{align}

\section{Conclusion}
\label{sec:Conclusion}
This paper derives the electromagnetic response for Gaussian fluctuations in the normal state of an ultraclean superconductor. 
The familiar Aslamazov-Larkin, Maki-Thompson, and Density of States diagrams are obtained, 
and an additional Gaussian-level diamagnetic diagram, not previously considered in the literature, also contributes to the electromagnetic response.
It is shown that this term is essential for establishing both gauge invariance and the absence of the Meissner effect in the normal state. 
The electrical conductivity was also considered. It was shown that the ultraclean MT, DOS, and AL diagrams sum to zero, and with the inclusion of the Gaussian-diamagnetic diagram the conductivity sum rule is satisfied.
This manuscript achieves a link between the functional-integral and diagrammatic-based approaches to superconductor fluctuation theory, 
showing lucidly how all the ultraclean fluctuation diagrams arise. As a result, the ultraclean fluctuation theory is now rendered gauge invariant and does not display a normal state Meissner effect.

\acknowledgments
The author thanks Brandon M. Anderson, Chun Chen, Pedro L. e S. Lopes, Hennadii Yerzhakov, Frank Marsiglio, and Joseph Maciejko for helpful discussions.
Thanks are also extended to Laura Fanfarillo for previously sharing their related work. 
Beneficial comments on the manuscript by Kathryn Levin and Andrey Varlamov are gratefully acknowledged. 
This research was supported by the Theoretical Physics Institute at the University of Alberta.

\appendix
\numberwithin{equation}{section}
\numberwithin{figure}{section}

\begin{widetext}
\appendix
\numberwithin{equation}{section}
\numberwithin{figure}{section}

\section{Aslamazov-Larkin diagram}
\label{sec:App_AL}
This excursus outlines the details in transforming from the position space fluctuation diagrams to their momentum space form. 
Specifically, starting from the position space definition of the fluctuation response, given in Eq.~$\left(\ref{eq:Action_Deriv}\right)$, 
it is shown how to obtain the momentum space diagrams in Eqs.~$\left(\ref{eq:AL_diagram}-\ref{eq:Gauss_diamag_diagram}\right)$. 
In position space the Aslamazov-Larkin diagram is defined by
\begin{equation}
\mathrm{AL}^{\mu\nu}\left(x,x^{\prime}\right) = -\int\prod_{i=1}^{4}dy_{i}L\left(y_{1},y_{2}\right)\Lambda^{\mu}\left(y_{2},x,y_{3}\right)L\left(y_{3},y_{4}\right)\Lambda^{\nu}\left(y_{4},x^{\prime},y_{1}\right).
\end{equation}
Converting this to momentum space then gives
\begin{eqnarray}
\mathrm{AL}^{\mu\nu}\left(x,x^{\prime}\right) & = & -\int\prod_{i=1}^{4}dy_{i}\prod_{j=1}^{6}\sum_{k_{j}}L\left(k_{1}\right)\Lambda^{\mu}\left(k_{2}+k_{3},k_{2}\right)L\left(k_{4}\right)\Lambda^{\nu}\left(k_{5}+k_{6},k_{5}\right)\nonumber \\
 &  & \times e^{ik_{1}\left(y_{1}-y_{2}\right)}e^{ik_{2}\left(y_{2}-y_{3}\right)+ik_{3}\left(y_{2}-x\right)}e^{ik_{4}\left(y_{3}-y_{4}\right)}e^{ik_{5}\left(y_{4}-y_{1}\right)+ik_{6}\left(y_{4}-x^{\prime}\right)},\nonumber \\
 & = & -\sum_{q}e^{-iq\left(x-x^{\prime}\right)}\sum_{p}L\left(p+q\right)\Lambda^{\mu}\left(p+q,p\right)L\left(p\right)\Lambda^{\nu}\left(p,p+q\right),\nonumber \\
 & \equiv & \sum_{q}e^{-iq\left(x-x^{\prime}\right)}\mathrm{AL}^{\mu\nu}\left(q\right).
\end{eqnarray}
Thus, in momentum space the Aslamazov-Larkin diagram is
\begin{equation}
\mathrm{AL}^{\mu\nu}\left(q\right) = -\sum_{p}L\left(p+q\right)\Lambda^{\mu}\left(p+q,p\right)L\left(p\right)\Lambda^{\nu}\left(p,p+q\right).
\end{equation}
The triangle vertex appearing in the Aslamazov-Larkin diagram is defined in Eq.~$\left(\ref{eq:AL_vertex}\right)$. Using the explicit form of the fluctuation propagator this becomes
\begin{eqnarray}
\Lambda^{\mu}\left(x,y,x^{\prime}\right) & = &-\left.\frac{\delta}{\delta A_{\mu}\left(y\right)}\left[G_{0}\left[A\right]\left(x,x^{\prime}\right)\widetilde{G}_{0}\left[A\right]\left(x^{\prime},x\right)+g^{-1}\delta\left(x-x^{\prime}\right)\right]\right|_{A=0},\nonumber \\
 & = & \left.2\int dx_{1}dx_{2}G_{0}\left(x,x_{1}\right)\frac{\delta G_{0}^{-1}\left[A\right]\left(x_{1},x_{2}\right)}{\delta A_{\mu}\left(y\right)}G_{0}\left(x_{2},x^{\prime}\right)\widetilde{G}_{0}\left(x^{\prime},x\right)\right|_{A=0},\nonumber \\
 & = & 2e\int dx_{1}dx_{2}G_{0}\left(x,x_{1}\right)\gamma^{\mu}\left(x_{1},y,x_{2}\right)G_{0}\left(x_{2},x^{\prime}\right)\widetilde{G}_{0}\left(x^{\prime},x\right).
\end{eqnarray}
The prefactor of two here arises due to particle-hole symmetry, namely, interchanging the particles and holes [and the appropriate coordinates] in the above expression produces the same result. 
Converting this to momentum space then gives
\begin{eqnarray}
\Lambda^{\mu}\left(x,y,x^{\prime}\right) & = & -2e\int\prod_{i=1}^{2}dx_{i}\prod_{j=1}^{5}\sum_{k_{j}}G_{0}\left(k_{1}\right)\gamma^{\mu}\left(k_{2}+k_{3},k_{2}\right)G_{0}\left(k_{4}\right)G_{0}\left(-k_{5}\right)\nonumber \\
 &  & \times e^{ik_{1}\left(x-x_{1}\right)}e^{ik_{2}\left(x_{1}-x_{2}\right)+ik_{3}\left(x_{1}-y\right)}e^{ik_{4}\left(x_{2}-x^{\prime}\right)}e^{ik_{5}\left(x^{\prime}-x\right)},\nonumber \\
 & = & -2e\sum_{p,q}e^{ip\left(x-x^{\prime}\right)+iq\left(x-y\right)}\sum_{k} G_{0}\left(k+q\right)\gamma^{\mu}\left(k+q,k\right)G_{0}\left(k\right)G_{0}\left(p-k\right),\nonumber \\
 & \equiv & \sum_{p,q}e^{ip\left(x-x^{\prime}\right)+iq\left(x-y\right)}\Lambda^{\mu}\left(p+q,p\right).
\end{eqnarray}
The last equality defines the momentum-space version of the triangle vertex, which is given by
\begin{equation}\label{eq:Lambda_vertex}
\Lambda^{\mu}\left(p+q,p\right) = -2e\sum_{k}G_{0}\left(k+q\right)\gamma^{\mu}\left(k+q,k\right)G_{0}\left(k\right)G_{0}\left(p-k\right).
\end{equation}
Thus, the complete expression for the Aslamazov-Larkin diagram is as given in Eq.~$\left(\ref{eq:AL_diagram}\right)$:
\begin{eqnarray}
\mathrm{AL}^{\mu\nu}\left(i\Omega_{m},\mathbf{q}\right) & = & -4e^{2}\sum_{k,l,p}G_{0}\left(l+q\right)\gamma^{\mu}\left(l+q,l\right)G_{0}\left(l\right)G_{0}\left(p-l\right)L\left(p+q\right)L\left(p\right)\nonumber \\
 &  & \times G_{0}\left(p-k\right)G_{0}\left(k\right)\gamma^{\nu}\left(k,k+q\right)G_{0}\left(k+q\right).
\end{eqnarray}
The minus sign here indicates that this diagram corresponds to bosonic transport; for example, a free bosonic correlation function has a minus sign in contradistinction to a free fermionic correlation function.
The prefactor of four also has a physical interpretation; writing the charge as $e^{*}=2e$ then shows that the AL diagram is just that of a free bosonic diagram, but with charge $2e$. 
The fermionic interactions manifest themselves in the triangle vertices and the explicit fluctuation propagator, but the physical understanding of the AL diagram can be found in that it is bosonic transport of fluctuating Cooper pairs.

\section{Maki-Thompson diagram}

In position space the Maki-Thompson diagram is defined by 
\begin{eqnarray}
\mathrm{MT}^{\mu\nu}\left(x,x^{\prime}\right) & = & -\left.2\int\prod_{i=1}^{6}dy_{i}L\left(y_{1},y_{2}\right)G_{0}\left(y_{2},y_{3}\right)\frac{\delta G_{0}^{-1}\left[A\right]\left(y_{3},y_{4}\right)}{\delta A_{\nu}\left(x^{\prime}\right)}G_{0}\left(y_{4},y_{1}\right)
\widetilde{G}_{0}\left(y_{1},y_{5}\right)\frac{\delta\widetilde{G}_{0}^{-1}\left[A\right]\left(y_{5},y_{6}\right)}{\delta A_{\mu}\left(x\right)}\widetilde{G}_{0}\left(y_{6},y_{2}\right)\right|_{A=0},\nonumber\\
 & = & -2e^{2}\int\prod_{i=1}^{6}dy_{i}L\left(y_{1},y_{2}\right)G_{0}\left(y_{2},y_{3}\right)\gamma^{\nu}\left(y_{3},x^{\prime},y_{4}\right)G_{0}\left(y_{4},y_{1}\right)
 \widetilde{G}_{0}\left(y_{1},y_{5}\right)\widetilde{\gamma}^{\mu}\left(y_{5},x,y_{6}\right)\widetilde{G}_{0}\left(y_{6},y_{2}\right).
\end{eqnarray}
Converting this to momentum space then gives
\begin{eqnarray}
&& \mathrm{MT}^{\mu\nu}\left(x,x^{\prime}\right)\nonumber\\
& = & 2e^{2}\int\prod_{i=1}^{6}dy_{i}\prod_{j=1}^{9}\sum_{k_{j}}L\left(k_{1}\right)G_{0}\left(k_{2}\right)\gamma^{\nu}\left(k_{3}+k_{4},k_{3}\right)G_{0}\left(k_{5}\right)G_{0}\left(-k_{6}\right)\gamma^{\mu}\left(-k_{7},-k_{7}-k_{8}\right)G_{0}\left(-k_{9}\right),\nonumber\\
& & \times e^{ik_{1}\left(y_{1}-y_{2}\right)}e^{ik_{2}\left(y_{2}-y_{3}\right)}e^{ik_{3}\left(y_{3}-y_{4}\right)+ik_{4}\left(y_{3}-x^{\prime}\right)}
e^{ik_{5}\left(y_{4}-y_{1}\right)}e^{ik_{6}\left(y_{1}-y_{5}\right)}e^{ik_{7}\left(y_{5}-y_{6}\right)+ik_{8}\left(y_{5}-x\right)}e^{ik_{9}\left(y_{6}-y_{2}\right)},\nonumber\\
& = & 2e^{2}\sum_{q}e^{-iq\left(x-x^{\prime}\right)}\sum_{k,p}L\left(p\right)G_{0}\left(k\right)\gamma^{\nu}\left(k,k+q\right)G_{0}\left(k+q\right)G_{0}\left(p-k-q\right)\gamma^{\mu}\left(p-k,p-k-q\right)G_{0}\left(p-k\right),\nonumber\\
& \equiv & \sum_{q}e^{-iq\left(x-x^{\prime}\right)}\mathrm{MT}^{\mu\nu}\left(q\right).
\end{eqnarray}
Thus, in momentum space the Maki-Thompson diagram is as given in Eq.~$\left(\ref{eq:MT_diagram}\right)$:
\begin{equation}
\mathrm{MT}^{\mu\nu}\left(i\Omega_{m},\mathbf{q}\right)=2e^{2}\sum_{k,p}G_{0}\left(p-k\right)\gamma^{\mu}\left(p-k,p-k-q\right)G_{0}\left(p-k-q\right)L\left(p\right)G_{0}\left(k\right)\gamma^{\nu}\left(k,k+q\right)G_{0}\left(k+q\right).
\end{equation}

\section{Density of States diagrams}

In position space the Density of States diagrams are defined by
\begin{eqnarray}
&&\mathrm{DOS}^{\mu\nu}\left(x,x^{\prime}\right)\nonumber\\
 & = & -2\left.\int\prod_{i=1}^{6}dy_{i}L\left(y_{1},y_{2}\right)G_{0}\left(y_{2},y_{3}\right)\frac{\delta G_{0}^{-1}\left[A\right]\left(y_{3},y_{4}\right)}{\delta A_{\mu}\left(x\right)}
 G_{0}\left(y_{4},y_{5}\right)\frac{\delta G_{0}^{-1}\left[A\right]\left(y_{5},y_{6}\right)}{\delta A_{\nu}\left(x^{\prime}\right)}G_{0}\left(y_{6},y_{1}\right)\widetilde{G}_{0}\left(y_{1},y_{2}\right)\right|_{A=0}\nonumber\\
 &  &- 2\left.\int\prod_{i=1}^{6}dy_{i}L\left(y_{1},y_{2}\right)G_{0}\left(y_{2},y_{3}\right)\frac{\delta G_{0}^{-1}\left[A\right]\left(y_{3},y_{4}\right)}{\delta A_{\nu}\left(x^{\prime}\right)}
 G_{0}\left(y_{4},y_{5}\right)\frac{\delta G_{0}^{-1}\left[A\right]\left(y_{5},y_{6}\right)}{\delta A_{\mu}\left(x\right)}G_{0}\left(y_{6},y_{1}\right)\widetilde{G}_{0}\left(y_{1},y_{2}\right)\right|_{A=0}.
\end{eqnarray}
Converting this to momentum space then gives
\begin{eqnarray}
 &  & \mathrm{DOS}^{\mu\nu}\left(x,x^{\prime}\right)\nonumber\\
 & = & 2e^{2}\int\prod_{i=1}^{6}dy_{i}\prod_{j=1}^{9}\sum_{k_{j}}L\left(k_{1}\right)G_{0}\left(k_{2}\right)\gamma^{\mu}\left(k_{3}+k_{4},k_{3}\right)G_{0}\left(k_{5}\right)\gamma^{\nu}\left(k_{6}+k_{7},k_{6}\right)G_{0}\left(k_{8}\right)G_{0}\left(-k_{9}\right)\nonumber\\
 &  & \times e^{ik_{1}\left(y_{1}-y_{2}\right)}e^{ik_{2}\left(y_{2}-y_{3}\right)}e^{ik_{3}\left(y_{3}-y_{4}\right)+ik_{4}\left(y_{3}-x\right)}e^{ik_{5}\left(y_{4}-y_{5}\right)}
 e^{ik_{6}\left(y_{5}-y_{6}\right)+ik_{7}\left(y_{5}-x^{\prime}\right)}e^{ik_{8}\left(y_{6}-y_{1}\right)}e^{ik_{9}\left(y_{1}-y_{2}\right)}\nonumber\\
 & + & 2e^{2}\int\prod_{i=1}^{6}dy_{i}\prod_{j=1}^{9}\sum_{k_{j}}L\left(k_{1}\right)G_{0}\left(k_{2}\right)\gamma^{\nu}\left(k_{3}+k_{4},k_{3}\right)G_{0}\left(k_{5}\right)\gamma^{\mu}\left(k_{6}+k_{7},k_{6}\right)G_{0}\left(k_{8}\right)G_{0}\left(-k_{9}\right)\nonumber\\
 &  & \times e^{ik_{1}\left(y_{1}-y_{2}\right)}e^{ik_{2}\left(y_{2}-y_{3}\right)}e^{ik_{3}\left(y_{3}-y_{4}\right)+ik_{4}\left(y_{3}-x^{\prime}\right)}e^{ik_{5}\left(y_{4}-y_{5}\right)}
 e^{ik_{6}\left(y_{5}-y_{6}\right)+ik_{7}\left(y_{5}-x\right)}e^{ik_{8}\left(y_{6}-y_{1}\right)}e^{ik_{9}\left(y_{1}-y_{2}\right)},\nonumber\\
 & = & 2e^{2}\sum_{q}e^{-iq\left(x-x^{\prime}\right)}\sum_{k,p}L\left(p\right)G_{0}\left(k+q\right)\gamma^{\mu}\left(k+q,k\right)
 G_{0}\left(k\right)\gamma^{\nu}\left(k,k+q\right)G_{0}\left(k+q\right)G_{0}\left(p-k-q\right)\nonumber\\
 & + & 2e^{2}\sum_{q}e^{-iq\left(x-x^{\prime}\right)}\sum_{k,p}L\left(p\right)G_{0}\left(k\right)\gamma^{\nu}\left(k,k+q\right)
 G_{0}\left(k+q\right)\gamma^{\mu}\left(k+q,k\right)G_{0}\left(k\right)G_{0}\left(p-k\right),\nonumber\\
 & \equiv & 2e^{2}\sum_{q}e^{-iq\left(x-x^{\prime}\right)}\mathrm{DOS}^{\mu\nu}\left(q\right).
\end{eqnarray}
Thus, in momentum space the Density of States diagrams are as given in Eq.~$\left(\ref{eq:DOS_diagrams}\right)$:
\begin{eqnarray*}
\mathrm{DOS}{}^{\mu\nu}\left(i\Omega_{m},\mathbf{q}\right) & = & 2e^{2}\sum_{k,p}G_{0}\left(k+q\right)\gamma^{\mu}\left(k+q,k\right)G_{0}\left(k\right)\gamma^{\nu}\left(k,k+q\right)G_{0}\left(k+q\right)L\left(p\right)G_{0}\left(p-k-q\right)\\
 & + & 2e^{2}\sum_{k,p}G_{0}\left(k+q\right)\gamma^{\mu}\left(k+q,k\right)G_{0}\left(k\right)L\left(p\right)G_{0}\left(p-k\right)G_{0}\left(k\right)\gamma^{\nu}\left(k,k+q\right).
\end{eqnarray*}

\section{Gaussian-level diamagnetic diagram}
\label{sec:App_GD}

This excursus outlines the details in obtaining Eq.~(\ref{eq:Gauss_diamag_diagram}). In position space the response is [see the last term in Eq.~(\ref{eq:Double_Deriv})]:
\begin{align}
\delta K^{\mu\nu}_{\mathrm{Fluc}}\left(x,x^{\prime}\right)&= 2\left.\int\prod_{i=1}^{4}dy_{i}L\left(y_{1},y_{2}\right)G_{0}\left(y_{2},y_{3}\right)\frac{\delta^{2}G_{0}^{-1}\left(y_{3},y_{4}\right)}{\delta A_{\mu}\left(x\right)\delta A_{\nu}\left(x^{\prime}\right)}G_{0}\left(y_{4},y_{1}\right)\widetilde{G}_{0}\left(y_{1},y_{2}\right)\right|_{A=0},\nonumber\\
 & =  -\frac{2e^{2}}{m}\delta^{\mu i}\delta^{\nu i}\int\prod_{i=1}^{4}dy_{i}L\left(y_{1},y_{2}\right)G_{0}\left(y_{2},y_{3}\right)\delta\left(y_{3}-x^{\prime}\right)\delta\left(y_{3}-y_{4}\right)\delta\left(x-x^{\prime}\right)
 G_{0}\left(y_{4},y_{1}\right)\widetilde{G}_{0}\left(y_{1},y_{2}\right).
 \end{align}
Converting this to momentum space then gives 
 \begin{eqnarray}
 \delta K^{\mu\nu}_{\mathrm{Fluc}}\left(x,x^{\prime}\right)
 & = & \delta\left(x-x^{\prime}\right)\frac{2e^{2}}{m}\delta^{\mu i}\delta^{\nu i}\int\prod_{i=1}^{2}dy_{i}\prod_{j=1}^{4}\sum_{k_{j}}L\left(k_{1}\right)G_{0}\left(k_{2}\right)G_{0}\left(k_{3}\right)G_{0}\left(-k_{4}\right)\nonumber\\
 &  & \times e^{ik_{1}\left(y_{1}-y_{2}\right)}e^{ik_{2}\left(y_{2}-x^{\prime}\right)}e^{ik_{3}\left(x^{\prime}-y_{1}\right)}e^{ik_{4}\left(y_{1}-y_{2}\right)},\nonumber\\
 & = & \delta\left(x-x^{\prime}\right)\frac{2e^{2}}{m}\delta^{\mu i}\delta^{\nu i}\sum_{k,p}L\left(p\right)G_{0}^{2}\left(k\right)G_{0}\left(p-k\right),\nonumber\\
 & \equiv & \sum_{q}e^{-iq\left(x-x^{\prime}\right)}\delta K^{\mu\nu}_{\mathrm{Fluc}}\left(q\right).
\end{eqnarray}
Thus, in momentum space the Gaussian-level diamagnetic diagram is as given in Eq.~(\ref{eq:Gauss_diamag_diagram}):
\begin{equation}
\delta K^{\mu\nu}_{\mathrm{Fluc}}\left(i\Omega_{m},\mathbf{q}\right) = \frac{2e^{2}}{m}\delta^{\mu i}\delta^{\nu i}\sum_{k,p}L\left(p\right)G_{0}\left(p-k\right)G_{0}^{2}\left(k\right).
\end{equation}
\end{widetext}

\section{Diamagnetic diagrams}
\label{sec:Diamag_diagrams}

This subsection shows that the diamagnetic diagrams can be expressed simply in terms of the derivative of the corresponding action with respect to the fermionic chemical potential.
The bare action is $S_{0}=-\mathrm{Trlog}\left(-\beta\mathcal{G}_{0}^{-1}\right)$,
and thus differentiating this with respect to the fermionic chemical potential $\mu$ then produces:
\begin{eqnarray}
\frac{1}{\beta V}\frac{\partial S_{0}}{\partial\mu}  =  -\sum_{p}\mathrm{tr}\left[\mathcal{G}_{0}\left(p\right)\tau_{z}\right]  =  -2\sum_{p}G_{0}\left(p\right).
\end{eqnarray}
Thus, the standard diamagnetic term is
\begin{equation}
\delta K_{0}^{\mu\nu}=-\delta^{\mu i}\delta^{\nu i}\frac{e^{2}}{m}\frac{1}{\beta V}\frac{\partial S_{0}}{\partial\mu}.
\end{equation}
The Gaussian-level diamagnetic term for the fluctuation response can be derived in a similar manner. 
The fluctuation action is $S_{\mathrm{Fluc}}=\mathrm{Tr\ log}\left(-gL^{-1}\right)$, and thus its derivative with respect to the fermionic chemical potential $\mu$ is:
\begin{eqnarray}
\frac{1}{\beta V}\frac{\partial S_{\mathrm{Fluc}}}{\partial\mu} & = & \sum_{p}L\left(p\right)\frac{\partial L^{-1}\left(p\right)}{\partial\mu},\nonumber \\
 & = & \sum_{p}L\left(p\right)\frac{\partial\Pi\left(p\right)}{\partial\mu},\nonumber \\
 & = & -2\sum_{k,p}L\left(p\right)G_{0}^{2}\left(k\right)G_{0}\left(p-k\right).
\end{eqnarray}
Here the definition of the pair susceptibility:
$\Pi\left(p\right)=\sum_{k}G_{0}\left(k\right)G_{0}\left(p-k\right)=L^{-1}\left(p\right)+g^{-1}$,
has been used. As a consequence of this result, the Gaussian-level diamagnetic diagram can be expressed as
\begin{equation}
\delta K_{\mathrm{Fluc}}^{\mu\nu}  =  -\delta^{\mu i}\delta^{\nu i}\frac{e^{2}}{m}\frac{1}{\beta V}\frac{\partial S_{\mathrm{Fluc}}}{\partial\mu}.
\end{equation}

Combining the above results, it follows that the full diamagnetic contribution is thus
\begin{equation}
\delta K_{0}^{\mu\nu}+\delta K_{\mathrm{Fluc}}^{\mu\nu} = -\delta^{\mu i}\delta^{\nu i}\frac{e^{2}}{m}\frac{1}{\beta V}\frac{\partial S_{\mathrm{Eff}}}{\partial\mu},
\end{equation}
where $S_{\mathrm{Eff}}=S_{0}+\mathrm{Tr\ log}\left(-gL^{-1}\right)$ is the effective action. 

\bibliography{Review_FluctuationPaper}

\end{document}